\documentclass[12pt]{iopart}

\usepackage{iopams}
\usepackage{graphicx}
\usepackage[ letterpaper,portrait, margin=1in]{geometry}
\begin{document}

\title{Spin-photon entanglement interfaces in silicon carbide defect centers}

\author{Sophia E. Economou}

\address{Department of Physics, Virginia Tech, Blacksburg VA 24061, USA}
\ead{economou@vt.edu}

\author{Pratibha Dev}

\address{Department of Physics and Astronomy, Howard University, Washington DC 20059, USA}

\vspace{10pt}

\begin{abstract}
Optically active spins in solid-state systems can be engineered to emit photons that are entangled with the spin in the solid. This allows for applications such as quantum communications, quantum key distribution, and distributed quantum computing. Recently, there has been a strong interest in silicon carbide defects, as they emit very close to the telecommunication wavelength, making them excellent candidates for long range quantum communications. In this work we develop explicit schemes for spin-photon entanglement in several SiC defects: the silicon monovacancy, the silicon divacancy, and the NV center in SiC. Distinct approaches are given for (i) single-photon and spin entanglement and (ii) the generation of long strings of entangled photons. The latter are known as cluster states and comprise a resource for measurement-based quantum information processing.
\end{abstract}

%Uncomment for keywords
\vspace{2pc}
\noindent{\it Keywords}: spin defect, silicon carbide, entanglement, quantum information processing, spin-photon entanglement, quantum communication
%\maketitle

\section{Introduction}

The capability to create entanglement between stationary qubits and flying qubits is of key importance in many quantum information processing protocols and technologies. Stationary-flying qubit entanglement allows for entanglement swapping or heralded entanglement, i.e., enables the ability to entangle systems that have never interacted directly. This in turn paves the way for technologies such as quantum communications, via the creation of quantum repeaters, as well as distributed quantum computing, over a chip or longer distances. In addition, it was proposed \cite{Lindner_PRL09,Economou_PRL10} that quantum emitters be used to generate long strings of entangled photons in a highly entangled graph state, known as a cluster state, which is a universal resource for quantum computing \cite{raussendorf_PRL01}. Cluster, and more generally graph states have also been proposed for all-photonic quantum repeaters \cite{Azuma_NatComm15}.

Optically active solid-state emitters are vigorously investigated for spin-photon entanglement generation. Exciting experiments have demonstrated spin-photon entanglement between a quantum-dot electron or hole spin and a photon \cite{DeGreve_nature12,Gao_nature12,Delteil_natphys16}, with Ref. \cite{Delteil_natphys16} achieving heralded entanglement of two hole spins through their emitted photons. Spin-photon entanglement has also been demonstrated between an NV center in diamond and an emitted photon \cite{Togan_nature11,Bernien_Nature13}, with an impressive recent experiment \cite{Hensen_Nature15} achieving heralded entanglement between two NV centers separated by more than 1 km and demonstrating for the first time loophole-free Bell inequality violation.

Over the last several years, there has been an effort in the community to explore defects beyond the NV center in diamond for quantum information processing \cite{Weber_PNAS10,Muller_NatComm14,Varley_PRB16}. A strong interest in silicon carbide defects has emerged, as they are hosted in an industrially mature, low-cost material and they emit at, or very close to, telecommunication wavelengths, making them very attractive candidates for long range quantum communications. Experimental demonstrations with these systems include a high-efficiency room-temperature single-photon source both at optical \cite{Castelletto2014} and microwave \cite{Kraus_NP14} frequencies. Spin-photon entanglement is still to be demonstrated with SiC defects.

In this work we develop explicit schemes for spin-photon entanglement in several SiC defects: the silicon monovacancy, the silicon divacancy, and the NV center in SiC. Distinct approaches are given for a single photon and spin entanglement and for the generation of long strings of entangled photons. The latter are known as cluster states and are a resource for measurement-based quantum information processing, including quantum computing \cite{raussendorf_PRL01} and all-photonic quantum repeaters \cite{Azuma_NatComm15}.
Density Functional Theory (DFT) is used to supplement the information regarding the defects themselves.

\section{Spin-photon entanglement}

\subsection{$\Lambda$-system}

\begin{figure}[ht]
\centering
\includegraphics[width=6.5 cm]{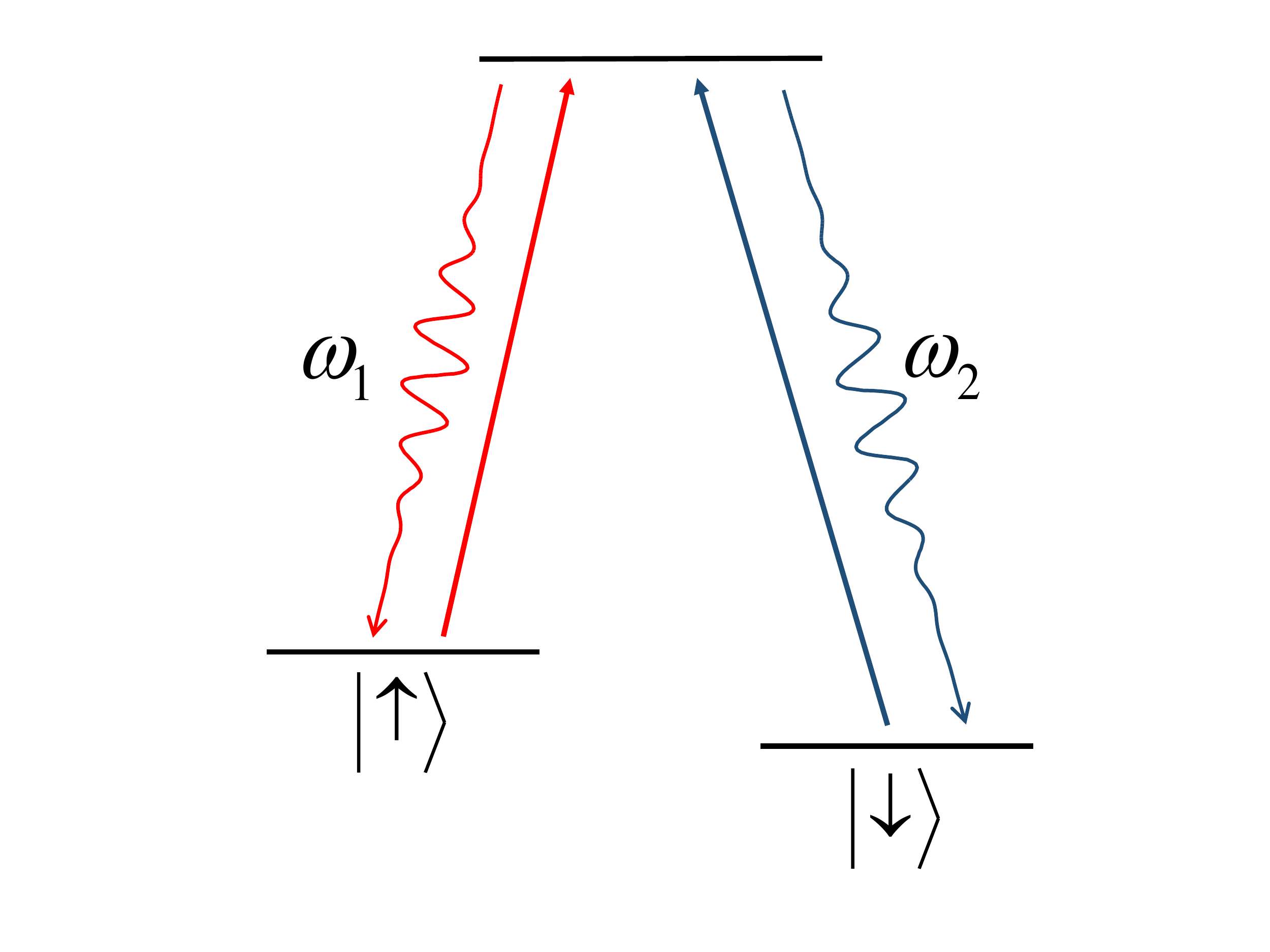}
\includegraphics[width=6.5 cm]{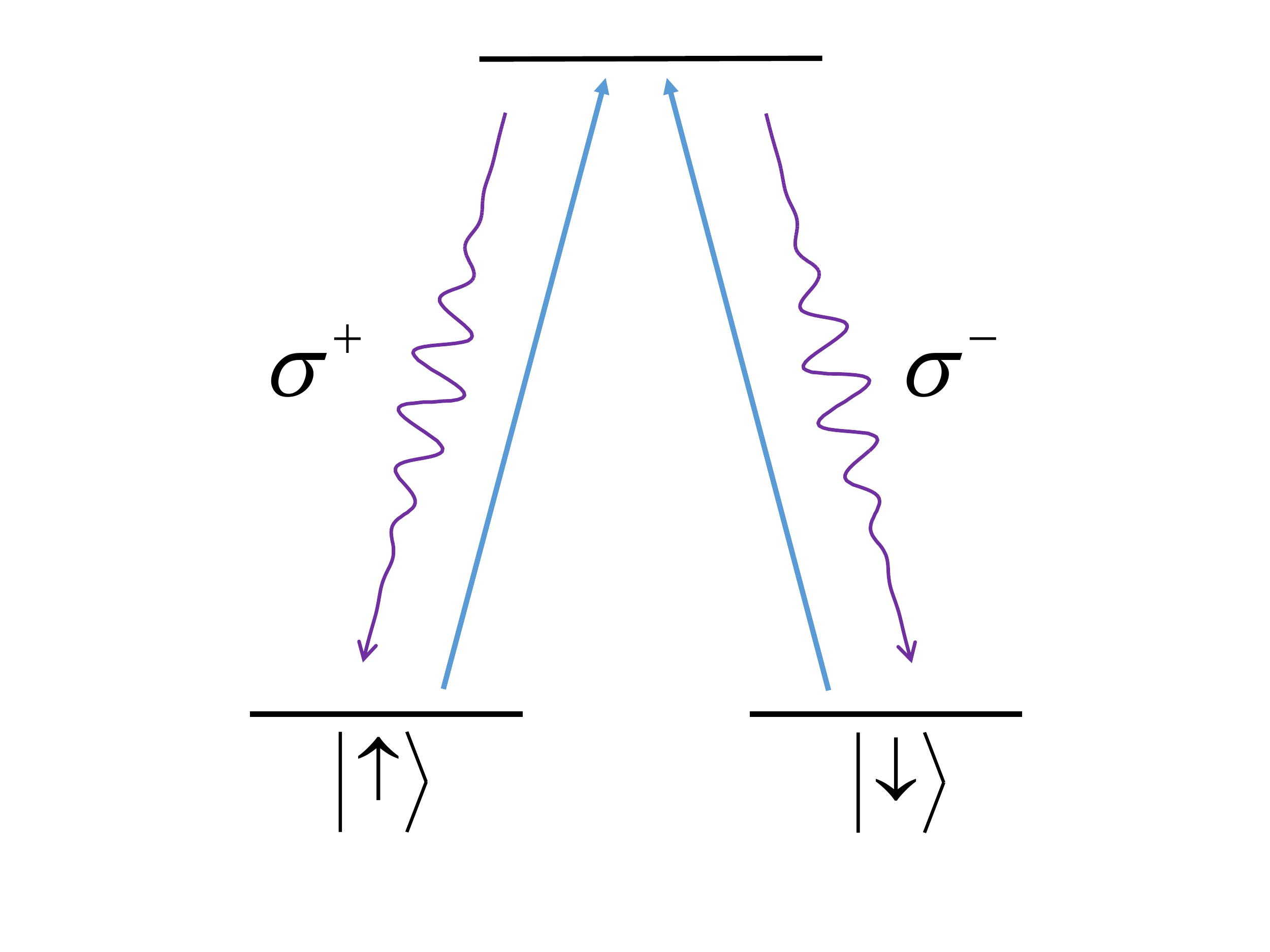}
\caption{Entanglement schemes for $\Lambda$ system.}
\label{GeneralLambda}
\end{figure}

The most common level structure to generate spin-photon entanglement is a three-level $\Lambda$ system. If the target entanglement is between the frequency of the emitted photon and the spin, the polarization of the two transitions should be the same. On the other hand, if the target entanglement is between the spin and the polarization degree of freedom of the photon, then the two transitions of the $\Lambda$ system should have equal frequencies. Fig. \ref{GeneralLambda} shows schematically these two possibilities.

\subsection{II-system}

\begin{figure}[ht]
\centering
\includegraphics[width=6.5 cm]{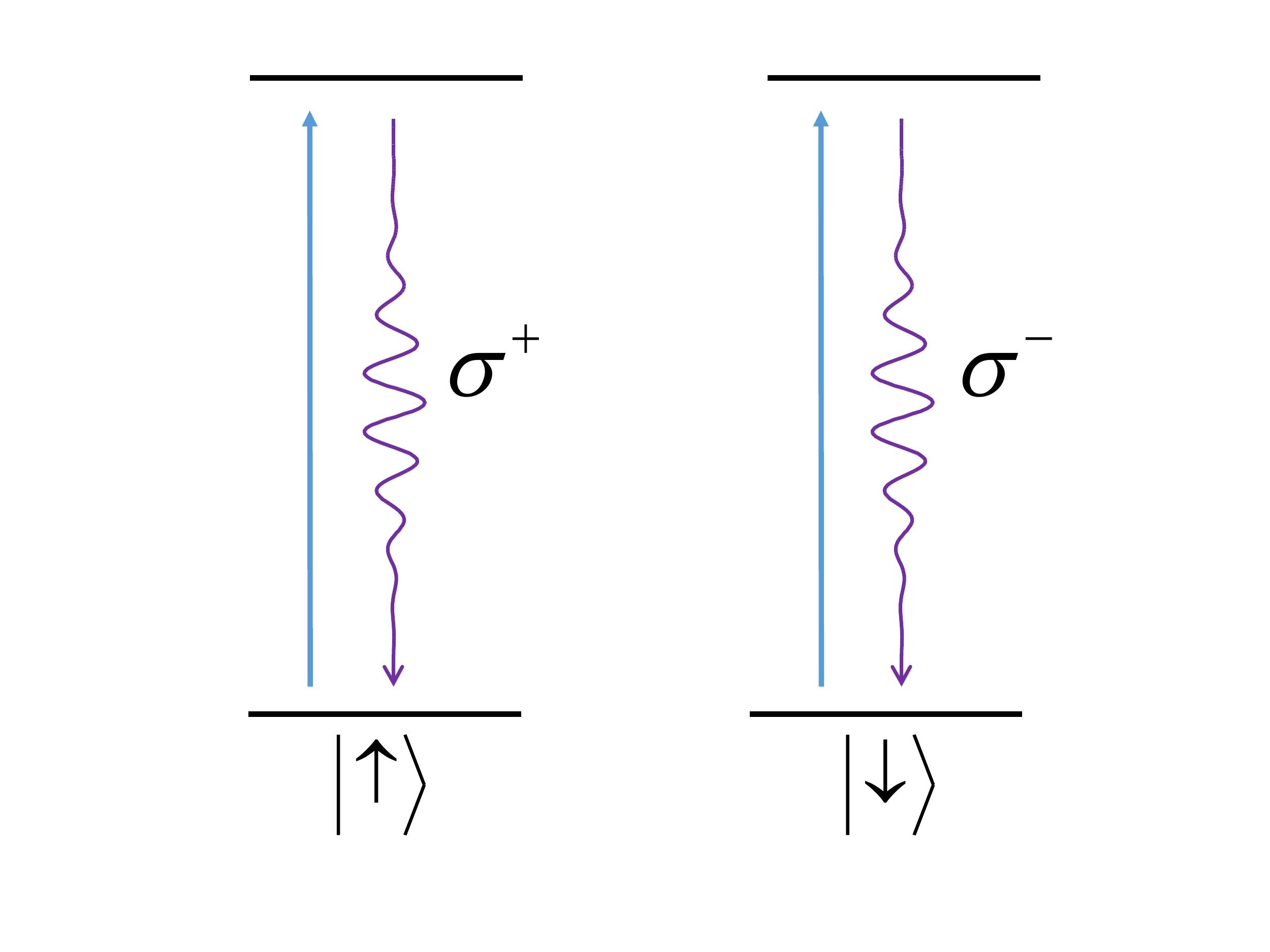}
\caption{Entanglement scheme for II system.}
\label{GeneralII}
\end{figure}

An alternative configuration of quantum levels for spin-photon entanglement generation is a four-level system, comprised of two two-level systems (defined here as the II-system). The two ground states and the two excited states are pairwise degenerate. In this approach, schematically shown in Fig. \ref{GeneralII}, the entanglement is between spin and photon polarization. Crucially, to generate entanglement, a superposition of the excited states has to be created. One of the difficulties in implementing this entanglement scheme is identifying systems for which all the necessary requirements (energy level structure, selection rules, ability to create superposition of excited states) exist simultaneously.

\section{Silicon Carbide defects}

Silicon carbide is a very interesting material. It comes in a variety of polytypes, some of which are important in industrial applications. It also hosts a very large number of possible defects, many of which are still uncharacterized. From the ones that are known and understood, the following have dominated proposals for applications in quantum information and have been studied experimentally the most in that context: the silicon-carbon divacancy ($\mathrm{V}_{\mathrm{Si}}-\mathrm{V}_{\mathrm{C}}$), the silicon vacancy ($\mathrm{V}_{\mathrm{Si}}$), and the NV center in SiC. Just like the NV-centers in diamond, these defects in SiC introduce atomic-like defect states that are spatially localized around the defect sites and carry a net spin. We used Density Functional Theory (DFT) to calculate these properties of the aforementioned defects. The Quantum-ESPRESSO package~\cite{QE-2009} was used to carry out the spin-polarized calculations within the generalized gradient approximation (GGA)~\cite{GGA} of Perdew-Burke-Ernzerhof (PBE)~\cite{PBE}. In order to highlight the localized nature of the defect wavefunctions, we created the defects at the h-site within a $6\times6\times2$ (576-atoms) supercell of 4H-SiC. The brillouin zone was sampled using a $\Gamma$-centered, $2\times2\times2$ k-point grid according to Monkhort-Pack method~\cite{kpoint2}. A different choice of polytype might change some details, but will leave most salient qualitative features of the defects unchanged. This is also true for the choice of h vs k site.

\subsection{Silicon-carbon divacancy}

\begin{figure}[ht]
\centering
\includegraphics[width=3.5 cm]{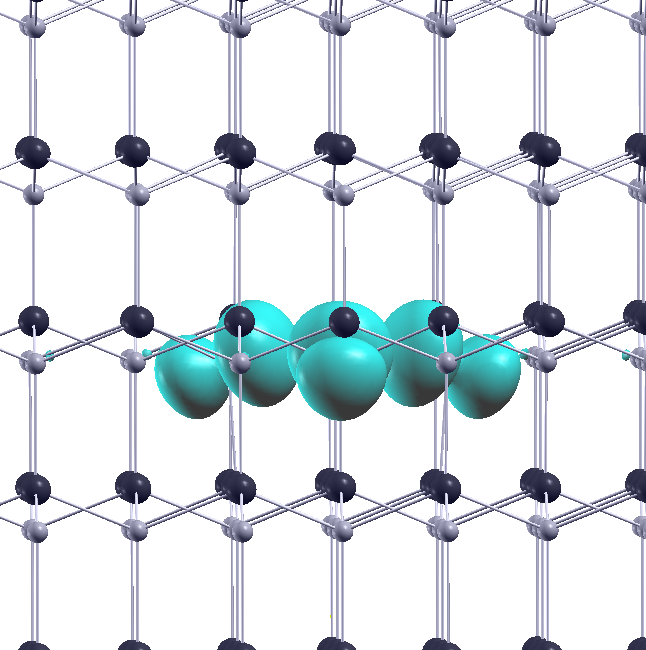}
\caption{Spin density plot, $\Delta \rho = \rho^{\uparrow} -\rho^{\downarrow}$ for the negatively charged $\mathrm{V}_{\mathrm{Si}}-\mathrm{V}_{\mathrm{C}}$ in SiC. The light grey spheres are carbons, the dark grey are silicon atoms.}
\label{VSiVC}
\end{figure}

The first defect we focus on is the silicon-carbon divacancy. As the name suggests, this defect is formed by two neighboring missing atoms (one Si, one C) in the SiC lattice \cite{Son_PRL06}. The symmetry of this (and of all defects discussed in this paper) is $C_{3v}$. The number of electrons associated with this defect is six, so that the divacancy is from a symmetry and an electron-number point of view identical to the NV center in diamond. Therefore, the energy levels and their symmetries are similar to those of the NV. Fig. \ref{VSiVC} shows the spin density isosurface ($\Delta \rho = \rho^{\uparrow} -\rho^{\downarrow}$) for the divacancy, showing the localization of the defect-induced spins onto the three nearest neighbor carbons. In recent years there have been important experimental demonstrations with this defect, including measurement of long coherence times even at room temperature \cite{Koehl_nature11}, control of isolated defects \cite{Christle_natmat14}, optical \cite{Falk_natcom13} and electrical \cite{Klimov_PRL14} control of the defect spin in various polytypes, as well as entanglement between electronic and nuclear spins \cite{Klimov_SciAdv15}.

\subsection{Silicon monovacancy}

\begin{figure}[ht]
\centering
\includegraphics[width=3.5 cm]{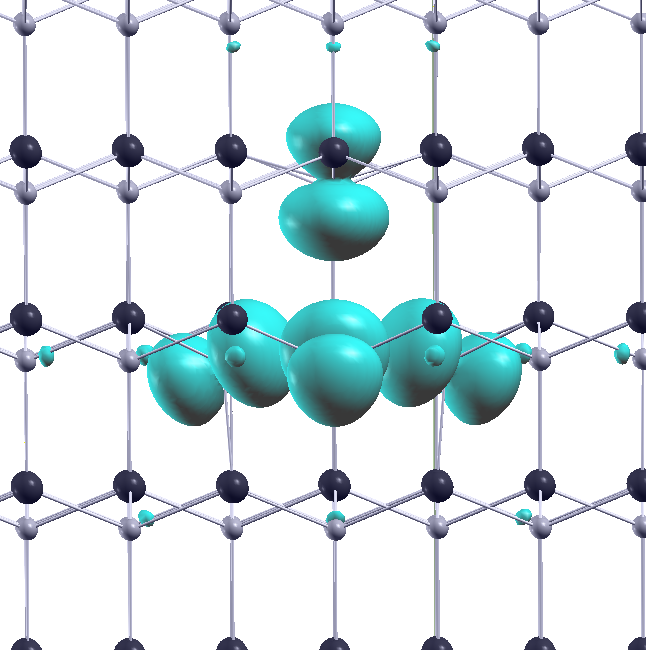}
\caption{Spin density plot, $\Delta \rho = \rho^{\uparrow} -\rho^{\downarrow}$ for the negatively charged $\mathrm{V}_{\mathrm{Si}}$ in SiC.}
\label{VSi}
\end{figure}

The negatively-charged silicon monovacancy is a defect consisting of a missing silicon atom, with the four neighboring carbons participating in the formation of the defect states. While this defect also has $C_{3v}$ symmetry, it is distinct from the other defects. The most important difference is that in its stable configuration it involves five active electrons (or, equivalently, three holes). Moreover, the distance of the carbon located along the C$_3$-symmetry axis from the vacancy, as calculated with DFT \cite{Soykal_PRB16}, is approximately equal to the distances of the basal carbons from the vacancy. As a result, the $C_{3v}$ symmetry of this defect is in fact very close to $T_d$ in the 4H and 6H polytypes \cite{janzen_physicab09,Soykal_PRB16}, and is exactly $T_d$ in 3C SiC. As a result of the odd number of active particles, $\mathrm{V}_{\mathrm{Si}}$ has a half-integer spin and due to the near $T_d$ symmetry of the defect, it is in particular a 3/2-spin system \cite{Mizuochi2002, Soykal_PRB16}. The four surrounding carbon atoms comprise to a very good approximation the extent of the defect wavefunction. This can be seen from the spin density isosurface plotted in Fig.\ref{VSi}.
%Moreover, the distance of the carbon located along the C$_3$-}symmetry axis from the vacancy, as calculated with DFT \cite{Soykal_PRB16}, is approximately equal to the distances of the basal carbons from the vacancy. As a result, the $C_{3v}$ symmetry of this defect is in fact very close to $T_d$ \cite{Soykal_PRB16}. The $\mathrm{V}_{\mathrm{Si}}$ defect has also been actively explored for quantum information applications.
This high-spin localized character of $V_{Si}$ defect has made it a promising candidate system for quantum information applications. Important experiments that are relevant to this work include the demonstration of optical and microwave control of these defects \cite{Riedel_PRL12}, including at the single-defect level \cite{Widmann_nmat15}, and their engineering as single photon emitters \cite{Fuchs_NatComm}.

\subsection{NV center}

\begin{figure}[ht]
\centering
\includegraphics[width=3.5 cm]{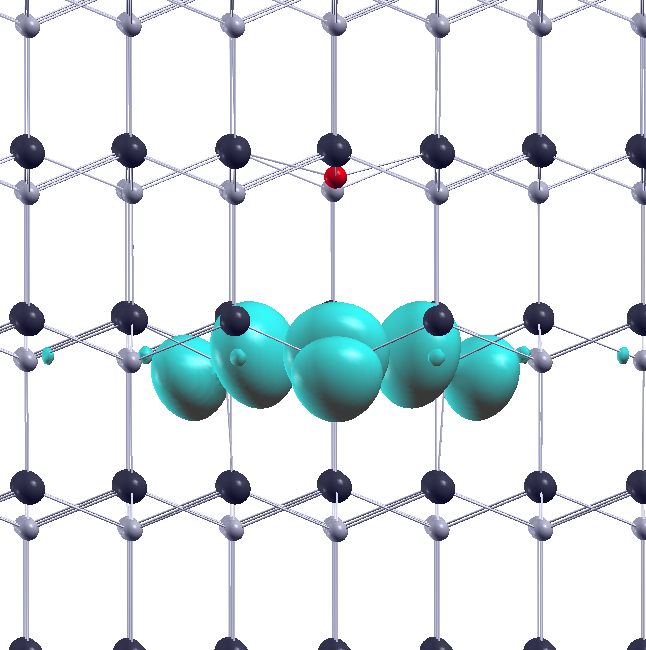}
\caption{Spin density plot, $\Delta \rho = \rho^{\uparrow} -\rho^{\downarrow}$, for the negatively charged NV-center in SiC. The red sphere is the nitrogen substituent.}
\label{NV}
\end{figure}

A less well-studied, but very attractive defect in SiC is the NV center, where, similarly to the NV in diamond, a substitutional nitrogen (${N}_{C}$) together with a silicon vacancy form a stable defect in the SiC lattice \cite{Muzafarova_MSF06,Pan_JAP10,vonBardeleben_PRB15,vonBardeleben_PRB16}. This defect is negatively charged, and has the exact same constituent atoms and active electrons as the NV in diamond. Therefore, qualitatively it has the same behavior in terms of energy levels, symmetries, and transitions, to NV-diamond. The NV center in SiC is also a deep defect. The isosurface plot of spin density in Fig.\ref{NV} shows that the net spin is localized on the three carbon atoms. The NV in silicon carbide combines attractive elements from its host lattice and the presence of the nitrogen: unlike the NV-diamond, it emits very close to the telecommunication wavelength, at 0.95 eV \cite{vonBardeleben_PRB15}. Just like the NV in diamond, it features the nuclear spin of the nitrogen, which can be strongly coupled to the electronic spin, in particular when the latter is optically excited, and can be used as a quantum memory. Because the NV center and the divacancy have the same electronic structure and selection rules, we will treat them below in a unified way, as the basic entanglement generation scheme will be the same in both cases.

\section{Spin-photon entanglement in divacancy and NV center}

\subsection{$\Lambda$-system}

Spin-photon polarization entanglement based on a $\Lambda$ system configuration in NV-diamond has been demonstrated experimentally \cite{Togan_nature11}. Since the divacancy and the NV-SiC have the same level structure as the NV in diamond, the entanglement protocol will work in essentially the same way. To create the $\Lambda$ system, a high strain regime is considered, where the multielectron states $E_x,E_y$ mix strongly into energy eigenstates of mixed orbital and spin character. The state $|E_+\rangle|{-}1\rangle + |E_-\rangle|1\rangle$ (where the first ket denotes the orbital degree of freedom and the second the spin) is selected as the excited state of the $\Lambda$ system. Initialization of the defect is carried out in the standard way, by optically illuminating the defect and taking advantage of the inter-system crossing, which populates the state $|0\rangle$.

\subsection{II-system}

\begin{figure}[ht]
\centering
\includegraphics[width=7.5 cm]{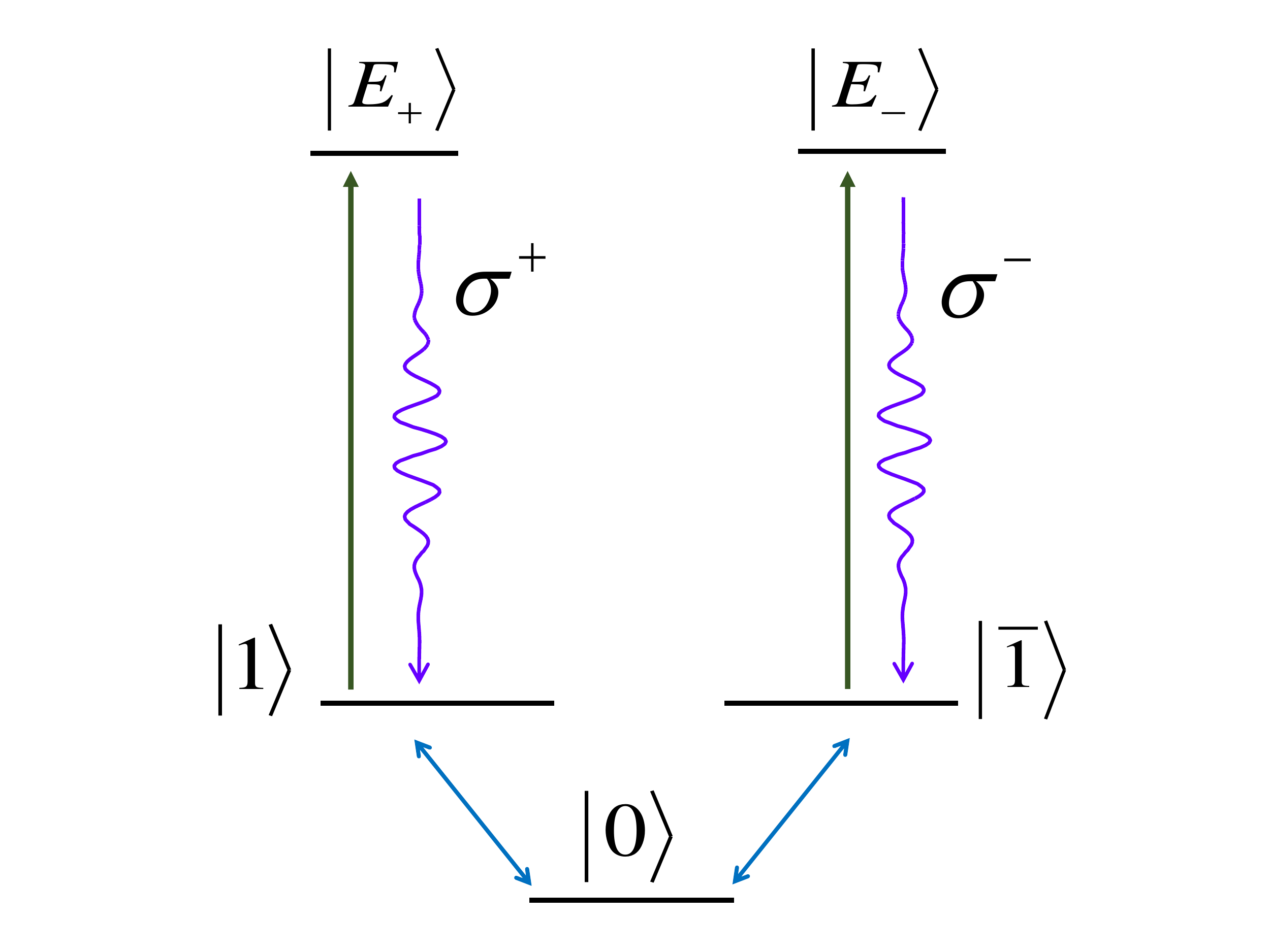}
\caption{Entanglement scheme for II system in divacancy and NV-SiC.}
\label{NVdivacancyII}
\end{figure}

An alternative physical regime is the low-strain regime, where states $E_+,E_-$ (or x,y) are degenerate but do not mix. In the absence of a magnetic field, the ground states $|\pm 1\rangle$ are also degenerate. The optical transitions between these two ground and excited states are circularly polarized, so that the cross transitions have $|\Delta m|=2$, and are forbidden due to selection rules. This way, a II-system can be formed, as shown in Fig. \ref{NVdivacancyII} (we define $|\overline{M}\rangle=|-M\rangle$). Even though the state $|0\rangle$ is not actively participating in the entanglement generation step (similarly to the $\Lambda$ system case), it is crucial in the overall protocol. \\

The entanglement protocol works in the following way: Initially, the system is initialized in state $|0\rangle$. A microwave $\pi$ pulse transfers the population from $|0\rangle$ to $|+\rangle \equiv |1\rangle+|-1\rangle$. A linearly polarized pulse transfers the population to state $|E_{+}\rangle|1\rangle + |E_{-}\rangle|-1\rangle$. Spontaneous emission of a photon results in the entangled state $|1\rangle|\sigma^+\rangle + |{-}1\rangle|\sigma^-\rangle$ (ignoring normalization for convenience). There are two ways to complete the protocol: one is simply to implement this step (excitation + spontaneous emission) $n$ times, such that a GHZ state $|1\rangle|\sigma^+\rangle ^n + |{-}1\rangle|\sigma^-\rangle ^n $ is generated. Alternatively, a more robust entangled state, a cluster state, can be formed by supplementing each `pump and collect' cycle with a microwave pulse that implements, in the $|\pm 1\rangle$ subspace, the unitary operation
\begin{eqnarray}
|1\rangle &\rightarrow & (|1\rangle + |{-}1\rangle)/\sqrt{2} \nonumber\\
|{-}1\rangle &\rightarrow & (|{-}1\rangle - |1\rangle)/\sqrt{2} .
\label{unitary}
\end{eqnarray}
This operation in matrix form in the $|\pm 1\rangle$ basis reads:
\begin{eqnarray}
U=\frac{1}{\sqrt{2}}
\left(\begin{tabular}{@{}ll} 1 & -1 \\ 1 & ~1 \end{tabular} \right).
\end{eqnarray}
The matrix $U$ is diagonal in the basis $|1\rangle\pm i |{-}1\rangle$ with eigenvalues $e^{\pm i\pi/4}$. Since an overall phase is immaterial, this matrix is equivalent to the diagonal matrix $(e^{ i\pi/2},1)$. Representing $U$ in this diagonal form in the superposition basis allows for a simple design of this operation. In particular, $U$ can be implemented through use of the auxiliary state $|0\rangle$. This works by applying a $2\pi$ pulse between states $|0\rangle$ and $(|1\rangle + |{-}1\rangle)$, with the appropriate detuning, such that a phase of $\pi/2$ is picked up after the cyclic evolution. This is generally a non-trivial requirement, as for a generic pulse shape, resonant pulses give a phase of $\pi$. To obtain the phase of $\pi/2$ (or any other arbitrary phase), microwave pulses of temporal envelope $\Omega(t)=\Omega_0 sech({\sigma t})$ and detuning $\Delta$ can be used. Based on the results in \cite{Economou_PRB06}, the desired phase of $\pi/2$ will be induced when $\Delta=\sigma$. Therefore, the cluster state protocol is realized in the following way: (i) initialize the system to state $|0\rangle$, (ii) use a microwave $\pi$ pulse to put the system in state $|1\rangle + |{-}1\rangle$, (iii) excite to state $|E_+\rangle|1\rangle + |E_{-}\rangle|-1\rangle$ using a resonant, linearly polarized laser, (iv) wait for spontaneous emission and collect the photon, (v) apply a detuned microwave pulse, as explained above, to implement $U$. Next, repeat steps (i)-(v) $n$ times to obtain a linear cluster state of $n$ photons. This chain will still be entangled with the defect spin, but it is simple to `cut' the chain by either measuring the defect spin state or the last emitted photon.

\section{Spin-photon entanglement in silicon monovacancy}

The monovacancy is qualitatively distinct from the defects discussed above. Due to the half integer spin of this defect, states come in pairs (Krammers' degeneracy). Thus, to split states of opposite $m_s$, a magnetic field should be used.

\subsection{$\Lambda$-system}

To form a $\Lambda$-system, the symmetry of the defect has to be lowered. In the case of the NV or divacancy, this can be done by nonzero strain. For the monovacancy, we proposed in Ref. \cite{Soykal_PRB16} to form a $\Lambda$ system via a magnetic field transverse to the symmetry axis of the defect. In this case, the transitions forming the $\Lambda$-system have the same polarization, but distinct frequencies. To see why this is the case, we consider the energy eigenstates under the transverse magnetic field. We assume that the B-field, pointing along $x$, is large enough, such that the Zeeman terms are much larger than the ZFS. Then we have
\begin{eqnarray}
|3/2,x\rangle &=& \frac{1}{2\sqrt{2}}|3/2\rangle + \frac{\sqrt{3}}{2\sqrt{2}}|1/2\rangle+\frac{\sqrt{3}}{2\sqrt{2}}|\overline{1}/2\rangle+\frac{1}{2\sqrt{2}}|\overline{3}/2\rangle \\
|1/2,x\rangle &=& -\frac{\sqrt{3}}{2\sqrt{2}}|3/2\rangle - \frac{1}{2\sqrt{2}}|1/2\rangle+\frac{1}{2\sqrt{2}}|\overline{1}/2\rangle+\frac{\sqrt{3}}{2\sqrt{2}}|\overline{3}/2\rangle \\
|\overline{1}/2,x\rangle &=& \frac{\sqrt{3}}{2\sqrt{2}}|3/2\rangle - \frac{1}{2\sqrt{2}}|1/2\rangle-\frac{1}{2\sqrt{2}}|\overline{1}/2\rangle+\frac{\sqrt{3}}{2\sqrt{2}}|\overline{3}/2\rangle \\
|\overline{3}/2,x\rangle &=& -\frac{1}{2\sqrt{2}}|3/2\rangle + \frac{\sqrt{3}}{2\sqrt{2}}|1/2\rangle-\frac{\sqrt{3}}{2\sqrt{2}}|\overline{1}/2\rangle+\frac{1}{2\sqrt{2}}|\overline{3}/2\rangle.
\end{eqnarray}

For a $\Lambda$-system to be realizable with appropriate pulses, we have to ensure there are no additional transitions excited that are not part of the $\Lambda$-system. If the ground and excited states are the same in terms of symmetries and quantum numbers, this will be an issue. Thus, the $A_2$ excited manifold \cite{Soykal_PRB16} cannot be used for this scheme, as both sets of states will be affected equally by the applied field. However, there exists an excited manifold of $E$ symmetry, which in addition features a nonzero spin-orbit splitting \cite{Soykal_PRB16}. As a result, the symmetry is lowered and a $\Lambda$-system is possible, because the large spin-orbit splitting prevents the transverse magnetic field from mixing states with the same orbital configuration and $|M_S|=3/2$. To assess the validity regime for this case, the Zeeman terms in the Hamiltonian should be much weaker than the spin-orbit. At the same time, we have assumed above that the B-field is strong compared to the ZFS. To find the appropriate parameter regime for the magnetic field quantitatively, we need an estimate both for the spin-orbit and for the spin-spin shifts. The spin-spin shift is known for the ground state to be about 70 MHz. Below we provide an estimate for the spin-orbit shifts, which due to symmetry are nonzero only in the excited states. More specifically, we are interested in the states $||e_\pm u v\rangle|\uparrow\uparrow\uparrow\rangle$, where the notation $||...\rangle$ denotes a Slater determinant, and the state has been expressed in terms of holes instead of electrons. We are interested in the spin-orbit shift of this state, i.e., the matrix element of the axial term $H_{SO}\equiv\sum_{j=1}^3 h_{SO}=\sum_{j=1}^3 \lambda_z \ell_{jz} s_{jz}$. Using the Slater-Condon rules and the symmetries of the single-particle states, the nonzero part of this matrix element can be written as
\begin{eqnarray}
\langle H_{SO} \rangle = Im\langle e_y|h_{SO}|e_x\rangle .
\end{eqnarray}
To proceed with a quantitative estimate of this term, it is helpful to notice that the molecular states $e_x, e_y$ are comprised only of (hybridized) atomic states belonging to the three basal carbons. That is, due to symmetry, the carbon along the symmetry axis does not participate in states $e_x, e_y$. Next, we write down the expression for the same part of the Hamiltonian in the case of the NV center in diamond. Again, using the Slater-Condon rules and symmetry, this matrix element can be expressed as $Im\langle e_y|h_{SO}|e_x\rangle$ . Using similar considerations as above, we see that the nitrogen states do not participate explicitly in the formation of states $e_x,e_y$. Thus, we can use the experimentally measured values from the NV center in diamond to obtain an estimate for the spin-orbit shifts in SiC defects. This estimate should be accurate up to corrections originating from the admixture of basis states beyond the three basal carbons, which will be different species in each case (silicon atoms mainly for SiC and carbon atoms for diamond). With these considerations, we estimate the spin-orbit shift (denoted as 3$\Delta$ in \cite{Soykal_PRB16}) to be $\Delta_{SO}\sim 5$ GHz \cite{Batalov_PRL09}. Finally, we note that there will be a further shift from the spin-spin interaction in the excited state (excited state ZFS). This was recently measured to be 410 MHz \cite{Simin_preprint15}, which is a relatively small correction to the SO shift, and thus will not change the qualitative behavior of the pinned $\pm 3/2$ projections in this excited state manifold.

Using our quantitative estimate of the spin-orbit, we restrict the B field to be such that $2 D\ll \omega_B \ll \Delta_{SO}\Rightarrow 70 MHz \ll \omega_B \ll 5 GHz $, which gives a B field on the order of 10 mT.

Now we turn to the question of spin-photon entanglement from this protocol.
The entanglement protocol works in the following way: First, the system is initialized via optical pumping and inter-system crossing to one of the ground states. More experiments are required, especially in the transverse B field regime, to determine the exact polarization process and outcome. Next, a pulsed laser excites the population to state $||uve_x+i e_y\rangle|\uparrow\uparrow\uparrow\rangle$. Because the excited state, having $M_S=3/2$, couples to all four ground states, the final state will be an entangled state, with the entangled degree of freedom of the photon being the frequency, with four terms, namely
\begin{eqnarray}
|\Psi\rangle = c_1 |3/2,x\rangle + c_2|1/2\rangle + c_2|\overline{1}/2\rangle +c_1 |\overline{3}/2,x\rangle.
\end{eqnarray}
The fact that this entangled state features all four ground states may make it useful for quantum information protocols on its own right. If, however, a qubit system is desired, we can exploit the fact that $c_1/c_2 =\sqrt{3}$ and filter out photons with frequencies outside the window corresponding to the transitions to the states $|3/2,x\rangle$ and $|\overline{3}/2,x\rangle$, without sacrificing too many of the emission processes.

\subsection{II-system}

Alternatively, the polarization of the photon can be used as the entangled degree of freedom in spin-photon entanglement or multi-photonic, cluster state entanglement. The II-system is formed by a pair of two lower states and a pair of two excited states. The lower states can either be states $|1/2\rangle$ and $|-1/2\rangle$ or $|3/2\rangle$ and $|-3/2\rangle$.

In the approach based on the two $|M_S|=1/2$ as the lower levels, the excited states are $||e_+uv\rangle(|\uparrow\uparrow\downarrow\rangle+|\uparrow\downarrow\uparrow\rangle+|\downarrow\uparrow\uparrow\rangle)/\sqrt{3}$ and $||e_-uv\rangle (|\downarrow\downarrow\uparrow\rangle+|\downarrow\uparrow\downarrow\rangle+|\uparrow\downarrow\downarrow\rangle)/\sqrt{3}$, where $|e_\pm\rangle \propto |e_x\rangle \pm i| e_y\rangle$. In terms of total angular momentum projection, the two transitions forming the II-system can be represented as $|1/2\rangle \longleftrightarrow |3/2\rangle$ and $|{-}1/2\rangle \longleftrightarrow |{-}3/2\rangle$. This system is reminiscent of transitions between electron and trion spin states in self-assembled quantum dots. Indeed, once the II-system is formed, the entanglement protocol is very similar to that developed for quantum dots \cite{Lindner_PRL09, Economou_PRL10}.

\begin{figure}[ht]
\centering
\includegraphics[width=8.5 cm]{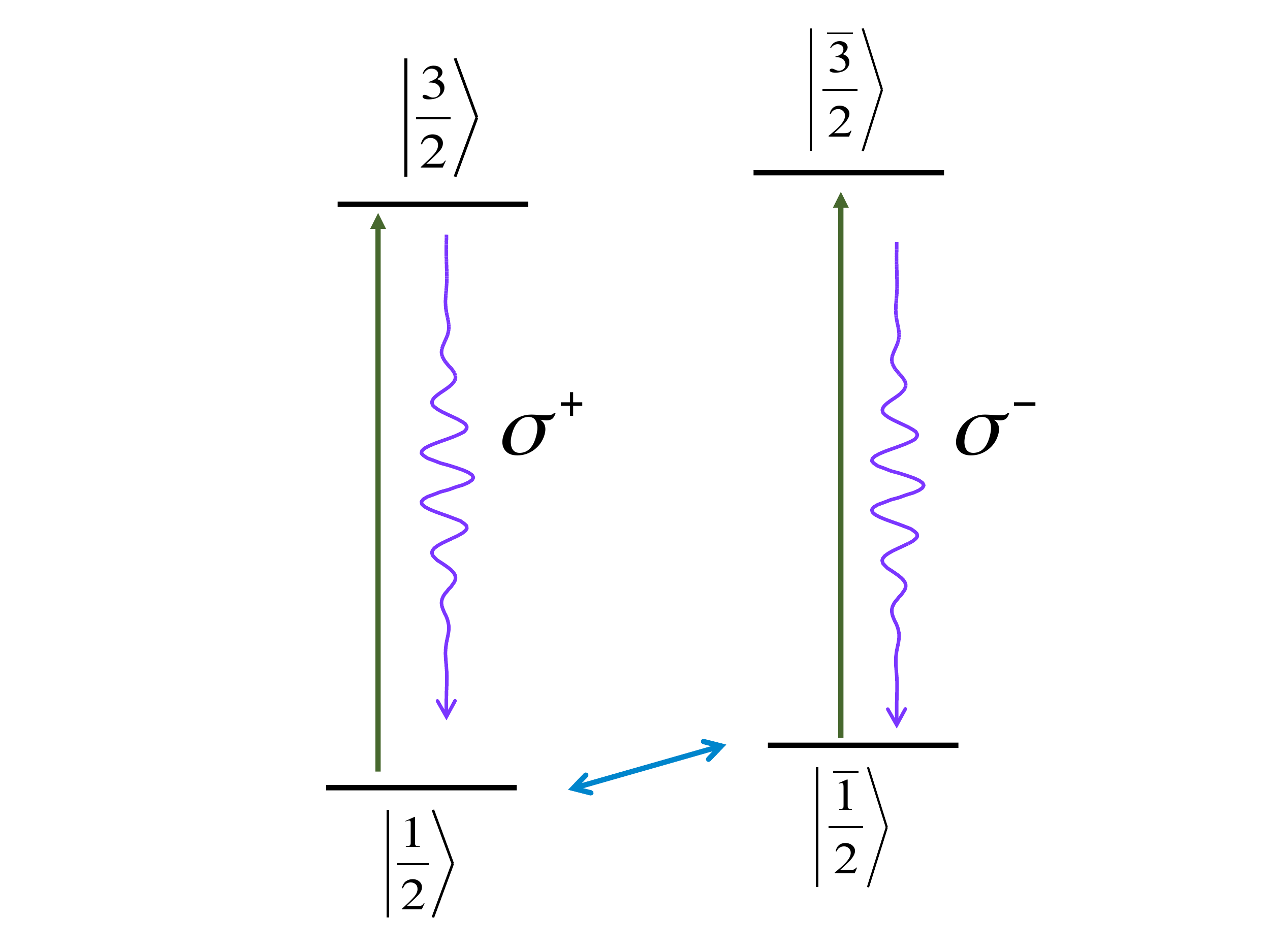}
\caption{Entanglement scheme for II system in Si vacancy based on the $|S_z|=1/2$ lower levels. The $|S_z|=3/2$ ground states are not shown because they are not involved in this scheme. }
\label{VSiIIa}
\end{figure}

The protocol proceeds as follows: the system is initialized in the state $|1/2\rangle + |{-}1/2\rangle$, and is subsequently excited by linearly polarized light to $|3/2\rangle + |{-}3/2\rangle$. This state decays spontaneously to the spin-photon entangled state $|1/2\rangle|\sigma^+\rangle  + |{-}1/2\rangle |\sigma^-\rangle $. Continuing this pump and collect protocol, just like above, gives a $n$-photon GHZ state. To obtain a cluster state, a Hadamard-like gate is needed in between the optical pumping pulses. We consider a weak magnetic field parallel to the symmetry axis of the defect, which splits the $M_S=\pm1/2$ states. The Hadamard gate is then implemented by a microwave pulse, resonant with the transition between states $|1/2\rangle$ and $|{-}1/2\rangle$, and with pulse area $\pi/2$.

The approach based on the $|S_z|=3/2$ lower levels operates very similarly: the II system is comprised of $|3/2\rangle$ and $|-3/2\rangle$ and the excited states $||e_+uv\rangle|\uparrow\uparrow\uparrow\rangle$ and $||e_-uv\rangle|\downarrow\downarrow\downarrow\rangle$. In terms of total angular momentum projection, the two transitions forming the II-system can be represented as $|3/2\rangle \longleftrightarrow |5/2\rangle$ and $|{-}3/2\rangle \longleftrightarrow |{-}5/2\rangle$.
\begin{figure}[ht]
\centering
\includegraphics[width=8.5 cm]{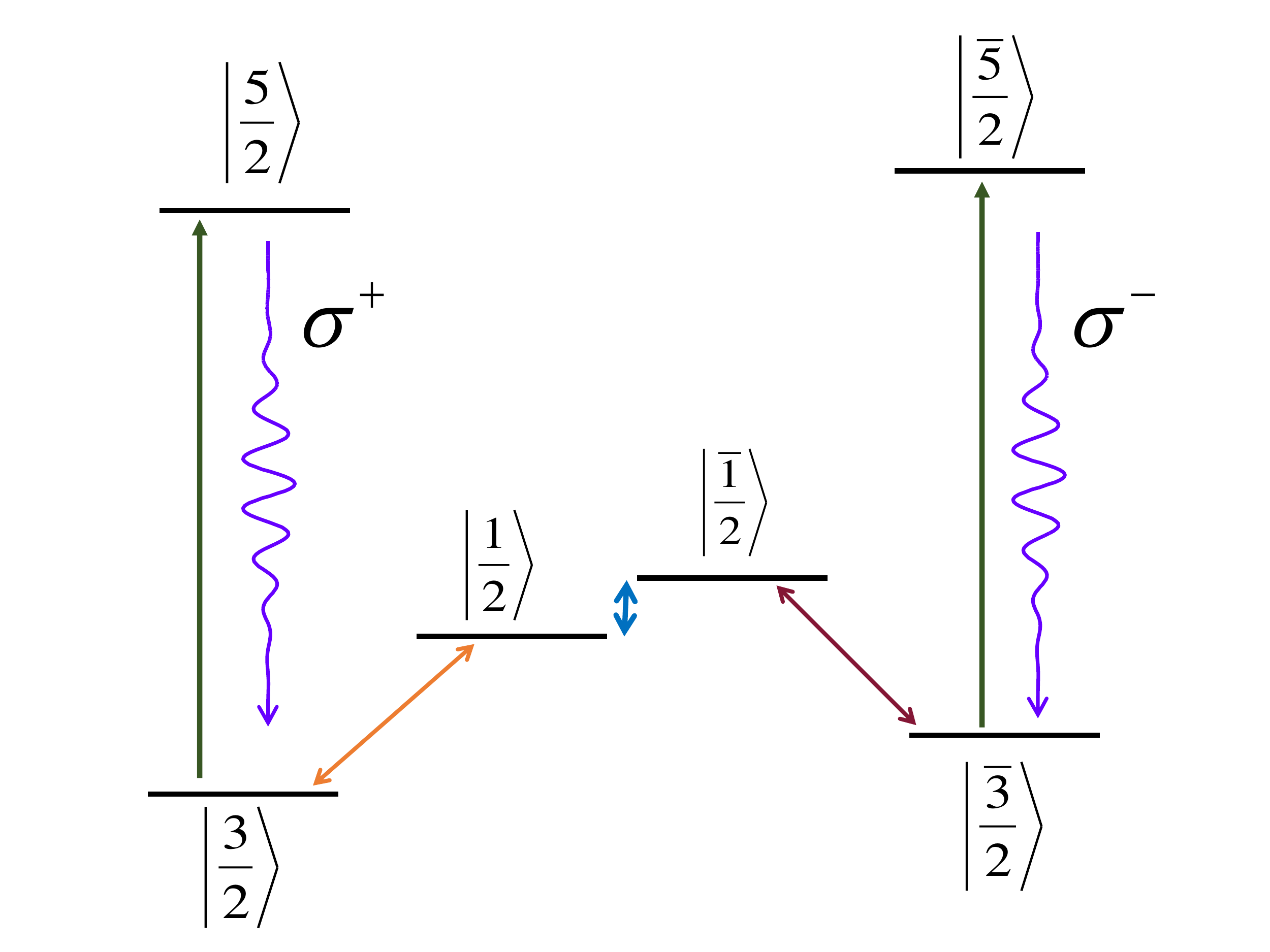}
\caption{Entanglement scheme for II system in Si vacancy based on the $|S_z|=3/2$ lower levels.}
\label{VSiIIb}
\end{figure}
Because the states with $|S_z|=3/2$ cannot be directly connected via external fields, the implementation of the Hadamard gate is more involved compared to the case above. To overcome this issue, we can make use of the $M_S=\pm1/2$ states. The gate can be carried out the following way: first, the population in $|3/2\rangle$ is transferred coherently to $|1/2\rangle$  and that in $|{-}3/2\rangle$ to $|{-}1/2\rangle$. Even though Fig. \ref{VSiIIb} shows these operations performed with two different microwave fields, in practice this should be possible with a single, broadband $\pi$ pulse. After this operation, a microwave pulse of the same type as in the $|M_S|=1/2$ case above is applied, i.e., resonant with the transition between states $|1/2\rangle$ and $|{-}1/2\rangle$, and with pulse area $\pi/2$. Following this pulse, a second $\pi$ pulse transfers the population back to the `qubit' states $|3/2\rangle$  and $|{-}3/2\rangle$. This completes one cycle of the protocol.

In our protocols, intersystem crossing transitions following the optical excitation should be avoided, as they would not produce an entangled photon. As silicon carbide defect research relating to quantum information applications is still at its infancy, many of the features of these defects are still not explored and understood fully, including details on intersystem crossing processes and their efficiency for the various defects. Different sites and different polytypes will most likely play a role in the efficacy of these protocols. For example, it has been shown that, unlike 4H-SiC, in 6H-SiC EPR spectra demonstrate a distinct behavior of a phase reversal at high optical intensities \cite{Astakhov_appmagres16}. The h vs k sites for a given polytype have also been shown to demonstrate a different behavior in terms of their photoluminescence intensity for circularly polarized transitions \cite{janzen_physicab09}. Given this complexity, more experimental and theoretical research is needed to identify which of our proposed schemes and which site and polytype will yield the highest fidelity. Use of cavities in these systems will also be crucial to the generation of high-fidelity spin-photon entanglement. Important recent work \cite{Bracher_preprint16} demonstrated the selective enhancement of two closely spaced defect zero-phonon lines in 4H-SiC through the Purcell effect, a result very relevant for spin-photon entanglement.

\section{Discussion and outlook}

In summary, we have provided detailed schemes for spin-photon entanglement interfaces based on silicon carbide defects. We have developed approaches for both $\Lambda-$ and II- systems and our proposals include entanglement schemes for either the polarization or the frequency degree of freedom. The variety of protocols we present in this work will allow experiments to adapt to the best behaved combination of defects, sites and transitions.

Our schemes can not only generate entanglement between the spin and a single photon, but also multiphotonic graph states. These are of key importance for measurement-based quantum information processing, including quantum computing and quantum communications.

Our proposed schemes are so far limited to entangled strings of photons (1D entanglement). To build up 2D entangled sheets, coupled defect centers can be used. The entanglement is first generated between the defect spins, and subsequent optical pumping and photon collection transforms this entanglement to photonic entanglement, similarly to Ref. \cite{Economou_PRL10}.  Recent experimental progress with positioning defects in a SiC sample \cite{Wang_preprint16} paves the path toward nearest-neighbor interactions between defects, and photonic crystal cavities \cite{Bracher_preprint16} could offer an alternative architecture that would allow long-range coupling between emitters. Determining the details of the physical interaction that can give the necessary coupling and devising the explicit 2D entanglement protocol are important open problems.

\section{Acknowledgements}
We would like to thank Cristian Bonato for useful conversations and for his suggestion regarding the spin-1/2 version of the II-system entanglement. 

\section*{References}
\bibliographystyle{iopart-num}
\bibliography{SiC}

\end{document}